\documentclass[prb]{revtex4}% Physical Review B

\usepackage{graphicx}% Include figure files
\usepackage{float}% Include figure files
\usepackage{dcolumn}% Align table columns on decimal point
\usepackage{bm}% bold math
\usepackage{verbatim}

\begin{document}
\preprint{APS/123-QED}

\title{Fabrication of microfluidic cavities using Si-to-glass anodic bonding }% Force line breaks with \\

%\author{N. Zhelev$^{1}$,  T. S. Abhilash$^{1}$, R. G. Bennett$^{1}$, E. N. Smith$^{1}$, B. Ilic $^{1}$, L. V. Levitin$^{2}$, X. Rojas$^{2}$, A. Casey$^{2}$, J. Saunders$^{2}$ and J. M. Parpia$^{1}$}

\author{N. Zhelev}
\affiliation{Department of Physics, Cornell University, Ithaca, NY, 14853 USA}
\author{T. S. Abhilash}
\affiliation{Department of Physics, Cornell University, Ithaca, NY, 14853 USA}
\author{R. G. Bennett}
\affiliation{Department of Physics, Cornell University, Ithaca, NY, 14853 USA}
\author{E. N. Smith}
\affiliation{Department of Physics, Cornell University, Ithaca, NY, 14853 USA}
\author{B. Ilic}
\affiliation{Department of Physics, Cornell University, Ithaca, NY, 14853 USA}
\author{L. V. Levitin}
\affiliation{Department of Physics, Royal Holloway University of London, Egham, TW20 0EX, Surrey, UK}
\author{X. Rojas}
\affiliation{Department of Physics, Royal Holloway University of London, Egham, TW20 0EX, Surrey, UK}
\author{A. Casey}
\affiliation{Department of Physics, Royal Holloway University of London, Egham, TW20 0EX, Surrey, UK}
\author{J. Saunders}
\affiliation{Department of Physics, Royal Holloway University of London, Egham, TW20 0EX, Surrey, UK}
\author{J. M. Parpia}
\affiliation{Department of Physics, Cornell University, Ithaca, NY, 14853 USA}

\date{\today}% It is always \today, today,

\begin{abstract}

We demonstrate the fabrication of $\sim$1.08  $\mu$m  deep microfluidic cavities with characteristic size as large as 7 mm $\times$ 11 mm or 11 mm diameter, using a silicon$-$glass anodic bonding technique that does not require posts to act as separators to define cavity height. Since the phase diagram of $^3$He is significantly altered under confinement, posts might act as pinning centers for phase boundaries. The previous generation of cavities relied on full wafer-bonding which is more prone to failure and requires dicing post-bonding, whereas the these cavities are made by bonding a pre-cut piece of Hoya SD-2 glass to a patterned piece of silicon in which the cavity is defined by etching. Anodic bonding was carried out at 425 $^{\circ}$C with 200 V, and we observe that pressurizing the cavity to failure ($>$ 30 bar pressure) results in glass breaking, rather than the glass-silicon bond separation. In this article, we discuss the detailed fabrication of the cavity, its edges, and details of the junction between the coin silver fill line and the silicon base of the cavity that enables a low internal-friction joint. This feature is important for mass coupling torsional oscillator experimental assays of the superfluid inertial contribution where a high quality factor ($Q$) improves frequency resolution. The surface preparation that yields well-characterized smooth surfaces to eliminate pinning sites, the use of transparent glass as a cover permitting optical access, low temperature capability and attachment of pressure-capable ports for fluid access may be features that are important in other applications.

\end{abstract}


% Glasses, acoustical properties of solids, vibrational states in disordered systems
                                   %PACS, the Physics and Astronomy
\maketitle

\section{Introduction}

It is now well established in experiments\cite{freeman,japanese,levitin1,levitin2,levitin3,zhelev} and on theoretical grounds \cite{barton1,barton2,ho,nagai,sauls,sauls2} that the phase diagram of the superfluid phases of $^3$He can be altered by confinement to sizes on the order of a few coherence lengths (77 nm to 14 nm as the pressure is varied from 0 bar to 34 bar). The zero temperature coherence length ($\xi_0$) is defined as $\hbar v_F/(2\pi k_BT_c)$, where $v_F$ is the $^3$He Fermi velocity, $T_c$ is the $^3$He superfluid transition temperature, and $k_B$ and $\hbar$ are Boltzmann's and Planck's constants. $^3$He displays the exotically paired superfluid/superconducting p-wave, {\it spin-triplet} state. Thus the observation of an altered phase diagram and the potential for realizing new phases under confinement in this system is of general interest.

Most experiments carried out to date have used multiple plates\cite{freeman,japanese} (with attendant inhomogeneities) to achieve confinement and also obtain the requisite signal to noise to carry out meaningful studies. The coherence length of $^3$He, $\xi_0$ that defines the relative confinement $D/\xi_0$ in a cell of height $D$ is strongly pressure dependent. To take advantage of this ability to vary the relative confinement, the cells should be able to withstand pressure. An alternative bonding scheme using silicon-to silicon bonding\cite{Rhee} includes pillars that may pin interfaces between competing phases.
In the recent past\cite{levitin1,levitin2,levitin3} experiments were described on the first generation of anodically bonded cavities that achieved good signal to noise with NMR. However, it was evident that scratches on the surfaces led to issues with pinning of interfaces between phases\cite{levitin2}. A secondary issue was interference of the signal from the vicinity of the fill line. In addition, there were difficulties associated with the point of attachment of the torsion rod/fill line to the silicon that had to be addressed since our previous scheme (described in Ref\cite{dimov}) had a significant failure rate at low temperatures as well as exhibiting a quality factor ($Q$) of only 70,000 at low temperatures\cite{dimov2}. A high {\it Q} or narrow linewidth is important for sensitive inertial resolution in a torsion pendulum. For macroscopic torsion pendulum experiments fabricated from coin silver, $Q$ factors on the order of 10$^6$ have been reported in the literature\cite{morely,casey,nyeki}. Coin silver is chosen (over beryllium copper) because of its weaker temperature dependent background\cite{morely}. Accordingly, we evolved changes in our process described in this paper that yield a {\it Q} factor in excess of $10^6$ while ensuring that the confined region be feature-free without supporting posts and scratches on the surface.

Attempts to study coherence-length thickness samples used films of $^3$He deposited on metallic surfaces. These yielded evidence of phase transitions \cite{crooker,seamus} and the observation of novel interfacial friction phenomena \cite{casey}. However, film experiments are necessarily limited to the vapor pressure and film thickness is difficult to control and calibrate.  Other complementary approaches to the study of confined $^3$He using regular geometries are detailed elsewhere\cite{davis,lee}. In the work described in Reference \cite{davis}, superfluid is confined to a precisely defined nanofluidic channel leading to a cavity, with the cavity and channel comprising a Helmholtz oscillator. The in-cavity capacitor measures the dielectric constant and transduces the pressure variation, while the mass in the channel oscillates in response to the pressure gradient. The cell is immersed in the fluid under study and thus does not have to be robust against pressure variations. The elimination of a pressure differential between confined region and the ``exterior" is also a feature of the devices pioneered by the Florida group\cite{lee} which incorporate a parallel plate geometry formed using the PolyMUMPs process \cite{lee2} to release a resonator in proximity to a solid wall allowing for transverse motion of the device. In both of these solutions, the presence of the bulk fluid would likely compromise the use of NMR where the signal from the bulk would overwhelm the signal from the cavity.

The anodic bonding technique described here\cite{Pomerantz,Anthony,Hui} is used in a variety of room temperature applications\cite{strook} incorporating microfluidic channels. One such example combines ultrasonics with microfluidic channels\cite{Nilsson}, and another \cite{quake} details integrated optical structures to visualize flows and mixing as well as flow of solids in liquids. The imaging of the drying of liquids in contact with specific porous geometries is another example of the use of anodic bonding and microfabricated fluidic chambers.\cite{vincent} The variations we demonstrate namely the design and attachment of pressure-capable fluidic ports, and surface preparation may be applicable in room temperature microfluidics.

The paper starts by describing the design of the cavities, outlines the process flow and cleaning steps used,  describes the patterning of the silicon, the design of the fill port and the polishing of the glass to eliminate scratches. The assembly jig for anodic bonding is described next along with the bonding process followed by the description of the metallization of the exterior of the cavity to achieve adequate thermalization. The surface roughness of both glass and silicon as measured using  an atomic force microscope (AFM) is then discussed together with the cavity depth measured using a profilometer scan and scanning electron microscope (SEM) imaging to characterize the edges of the cavity.  Finally details of assembly of the fill line are provided followed by results of pressure testing and finite element simulations on a cavity identical to that used for NMR imaging experiments.

\section{Experimental Details: Design and Cell Fabrication}

\subsection{Cavity Designs}

Wafer handling equipment at the Cornell NanoScale Science and Technology Facility (CNF) had significant difficulty in accommodating 3 mm thick wafers that we had deployed in the first generation devices\cite{dimov}. Thus we focussed on a design that would allow us to use 1 mm thick silicon and glass for both nuclear magnetic resonance (NMR) and torsion oscillator (TO)\cite{zhelev} experiments.

Figure 1 shows the mask design/cell pattern on each wafer. The grey color region shows the $\sim$1.08 $\mu$m recess for the cavity. Two types of devices were fabricated, one for torsion oscillator (TO) experiments (Fig. 1a) and other for NMR experiments (Fig. 1b). In both designs the devices are defined by an etched-through trench (blue in figure 1). The devices are connected to the wafer by thin tabs that can easily be broken to release the device from the wafer after processing. For the TO, each cut out terminated in a pair of tabs that could be broken to release the device from the surrounding silicon wafer. For the NMR cells the rectangular pattern had tabs at each corner that could be broken to release the devices. We started with a double side polished 100 mm diameter silicon wafer with 1 mm thickness and crystal orientation [100].

In the past\cite{levitin1,levitin2}, NMR experiments were able to resolve the fluid signal from a nominal height of 680 nm cavity embedded in 3 mm thick silicon and glass. Thus we sought to make a cavity design that had the same cavity shape (11 mm $\times$ 7 mm). However, to minimize bowing of the cavity under pressure, in the majority of the new cavities we introduced a 1 mm wide  ``septum" in the center of the cavity (Fig. 1b). We also incorporated (in some cavities) a second symmetric fill port to allow the possibility of introducing dc and ac flow through the cavity or a bulk sample located remotely from the fill line. Each wafer had either 16 identical TO devices patterned into it or 20 NMR cells, of which 11 had the septum, 5 had no septum, and 4 had two fill holes.

For the torsion pendulum, we sought to achieve a fluid to pendulum inertia ratio of $\approx$ 3 parts in 10$^6$. To obtain an inertia  resolution of 1.5 parts in 1000 we would require a frequency resolution of 1 part in 10$^9$ which
required the quality factor of our pendulum $> 10^6$ and also a reduction of mechanical and electrical noise in our setup. Our target dimension (cavity height) for these series of experiments was 1.1 $\mu$m, and the thickness of silicon and glass were each 1 mm. The design that evolved was a cylindrical cavity with a diameter of 11 mm, to be embedded in a silicon sample of 14 mm diameter. The fill line - torsion oscillator attachment point is located at the center of the cavity and consists of a 335 $\mu$m through hole. To reduce bowing and prevent collapse of the cavity during bonding, we incorporated a C-shaped support region centered in the cell with an outer diameter of 4.00 mm and inner diameter of 1.50 mm. Thus, the cavity that dominates the superfluid signal (main experimental cavity) has an annular shape of inner diameter of 4.00 mm and outer diameter of 11.00 mm.  The central region encompassing the fill line was connected to the main experimental cavity through a channel of 1.25 mm length and 0.60 mm width. Both the ``C" shaped region and the periphery are designed to be bonded to the glass lid.

The fluid inertia can be expressed as $I_f = (1/2)\pi(r_o^4-r_i^4) h \rho$, where $r_o, r_i$ are the outer and inner radii of the cavity (5.50 mm and 2.00 mm respectively), $h$ is the height (1.08 $\mu$m), and $\rho$ is the density of the $^3$He. At vapor (effectively zero) pressure, the density of $^3$He fluid is 0.0819 g/cm$^3$, yielding a fluid inertia of ~1.25 $\times$ 10$^{-6}$ g-cm$^2$. With a density of 2.33 g/cm$^3$, the corresponding silicon inertia is 8.78 $\times$ 10$^{-2}$ g-cm$^2$, and that of the octagonal glass lid (density 2.60 g/cm$^3$, 14 mm across flats) is 1.09 $\times$ 10$^{-1}$g-cm$^2$. Thus the expected fractional frequency shift on account of the $^3$He, $\Delta f/f$ = 1/2 $I_f$/$I_s$ = 3.17 $\times$ 10$^{-6}$. Our expected frequency shift on filling was 51.5 mHz/(g-cm$^{-3}$) and the measured value was 46.1 mHz/(g-cm$^{-3}$) likely due to the misalignment of the lid and base in assembly. The nominal resonant frequency was 1329.15 Hz.

\begin{figure}[H]
\begin{center}
    \includegraphics [angle=0,width=12cm]{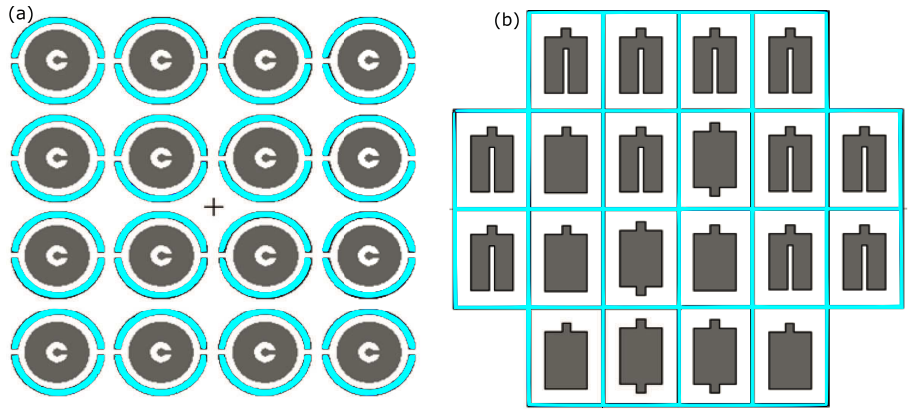}
    \caption{ An overview of the pattern on each silicon wafer. Grey layer is the region of the cavity space where helium will reside, the blue region is etched through the wafer. Not shown at this scale is the through hole for the fill line and the backside concentric circles. a) Wafer layout containing cells for torsion pendulum experiments. Each cell is spaced 18.3 mm center-to-center horizontally to fit into a standard 100 mm wafer. Each cell has a pair of ``tabs", narrow stubs of silicon to allow individual discs to be broken off from the parent wafer for processing.   b) Wafer design for NMR experiments. Adjacent NMR cells are spaced 10.4 mm apart horizontally so that the 20 cell pattern fits into a standard 100 mm diameter wafer. The ``tabs" holding the NMR cells to the parent wafer are not shown.}
    \label{figure1}
\end{center}
    \end{figure}

\subsection{The process flow to fabricate a cavity}

The fabrication process can be divided into five major steps. First, define and create the step in the silicon surface for the cavity. Second, etch through the fill line hole and the backside concentric circles to enable the attachment of the fill line to the silicon with epoxy. Third, polish and cut the glass. Fourth, clean and prepare the surfaces, and bond the silicon and glass to complete the cell. The final step is to deposit a thick film of silver to help thermalize the silicon and glass at ultra-low temperatures. Figure 2 shows the schematic of the process flow.

\begin{figure}[H]
\begin{center}
    \includegraphics [angle=0,width=12cm]{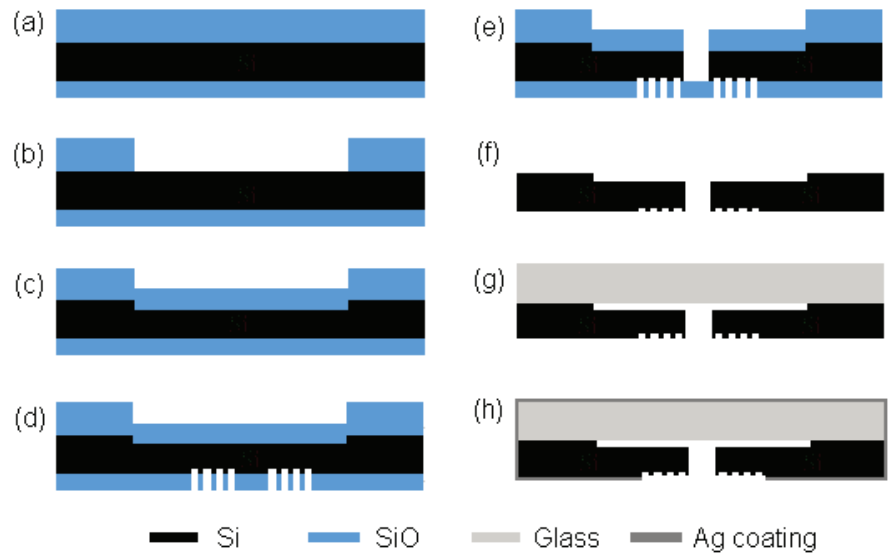}
    \caption{ Steps used in the fabrication 1.08 $\mu$m deep cavity in Si. (a) Grow thick thermal oxide followed by deposition of an additional 2 $\mu$m of PECVD oxide. (b) Etch the oxide in the cavity region (Dry + Wet). (c) Thermally oxidize the wafer to define cavity in Si. (d) Using DRIE, etch a series of concentric circles around the fill line. (e) Etch using DRIE through the wafer to define the fill line hole. (f) Remove all oxide using 49\% HF. (g) Bond Si to glass piece (h) Sputter deposit 1 $\mu$m silver film with $\sim$10 nm Ti adhesion layer (TO cell) on the outside surfaces. The silver coating did not extend around the sides of the NMR cell. Reproduced with permission (adapted from an earlier version of the figure) from Nature Communications, {\bf 8} 15963 (2017). Copyright 2017 Nature Publications under CC BY 4.0.}
    \label{figure2}
\end{center}
    \end{figure}
\noindent\\

\subsection{Cleaning procedures}
Cleaning was carried out before each lithography step and before oxidation steps used to define the cavity dimensions. Additional cleaning was carried our prior to assembly. The RCA SC-1 clean recipe follows the procedure listed here: 1 part NH$_{4}$OH (10-35\% NH$_3$), 1 part H$_{2}$O$_{2}$ (29-32$\%$ solution), and 5 parts of deionized water were heated to $~$75-80$^{\circ}$C. The solution was stirred vigorously using a magnetic stirrer while the pieces were immersed. The pieces were immersed in the stirred solution for 10 minutes between each lithography or oxidation step and for about half an hour before final assembly for bonding. The RCA SC-2 clean recipe  was performed with a solution of 6 parts of deionized water, 1 part HCl (33-40$\%$ solution), and 1 part of H$_2$O$_2$ (29-32$\%$ solution) at 70$^{\circ}$C, for 10 minutes.

\subsection{Defining the cavity}

A modified local oxidation process is used to define the cavity to confine the $^3$He. In the process, the silicon in the cavity region is converted to silicon oxide to a depth greater than the surrounding area, producing a recess in the silicon once all the oxide is removed. We could have etched the silicon surface using Reactive Ion Etching (RIE); however, this would have produced a much rougher finish at the bottom surface of the cavity and also would have resulted in significant variations of the cavity depth across the wafer and possibly also across a single cell. Cavity depth control and surface smoothness are extremely important for the experiments\cite{zhelev}.

To prepare the wafer for patterning we grew a thick oxide (2.5 to 3.5 $\mu$m) on both sides of the wafer by exposing the wafers for 12 hours at 1200 $^{\circ}$C in an atmospheric furnace. Hydrogen and oxygen (no HCl) were flowed into the furnace tube, providing the necessary conditions for wet oxidation. As the film thickness increases, the rate of the thermal oxide growth slows down significantly. Thus to get even thicker films, we needed to deposit extra oxide rather than growing more. A Plasma Enhanced Chemical Vapor Deposition (PECVD) process using SiH$_4$ and N$_2$O as precursors was used to deposit additional 2 $\mu$m of oxide on the front wafer surface. PECVD deposition is fairly fast, but the film quality is significantly lower, displaying lower density and large amount of incorporated hydrogen in the PECVD film compared to the thermally grown film. However, since we use this film as a mask layer, this is not as important and using PECVD oxide is acceptable. However, the lower density of the oxide layer affects the calculation of diffusion growth of oxide (discussed later in this section), and may compromise the accuracy of the desired depth so the PECVD film is only used as an additional masking layer. At the end of this step the wafer is at the point illustrated in Fig. 2a.

The next step was to spin, bake and develop positive photoresist (SPR 220-3.5) following a standard procedure. The cavity pattern was exposed using UV light, and the regions of the resist exposed to it were dissolved away in the developer solution (AZ 726, 2.38\% TMAH in H$_2$O). Since we were using a very thick (3-4 $\mu$m) photoresist, a proximity plate bake was used (115 $^{\circ}$C for 120 s). The wafer was not quenched on a cold surface to avoid cracking the resist and instead left to cool by itself in the wafer box. The wafers were then placed in a Reactive Ion Etch (RIE) chamber where a CHF$_3$ + O$_2$ recipe was used to dry etch and so remove the majority of the oxide in the cavity region. The last 150-200 nm of oxide was etched away using buffered oxide etch (BOE 1:6) since the plasma etch also attacks the Si. Because the BOE does not etch the Silicon, the etch is terminated exactly at the SiO$_2$-Si interface and thus the BOE leaves the silicon surface smooth whereas the RIE would leave it rough. At the end of this step the wafer is at the point illustrated in Fig. 2b.

After removing all the resist and carefully cleaning the wafers, the wafers were placed back into the oxidation furnace for a second oxidation step. The rate of oxide growth is given by the Deal-Grove model\cite{deal-grove,Tables}. For very thin oxide films the film thickness grows linearly with time, while for thick films the film thickness grows with the square root of the time. The silicon surface outside the cavity region is covered by a very thick layer of oxide (more than 2.5 $\mu$m). The rate of growth of oxide in this (thick oxide) region will be much slower than that in the cavity region where the silicon surface is exposed. A further point to account for is that for every 1 nm of silicon that is ``consumed", 2.25 nm of depth of oxide is produced. Thus, to create a step of $\times$ nm, we need to create a difference in the oxide growth inside and outside the cavity of 2.25$\times$. At the end of this step the wafer is at the point illustrated in Fig. 2c.

\subsection{The fill-line port}

The fill line port should have straight walls and a well-defined diameter. Previous attempts to drill the hole using diamond drills or grind a hole using diamond slurry \cite{dimovthesis}  proved possible but extremely difficult. Fear of contamination and of scratching the working surfaces was another reason why we chose to make the hole using the tools inside the CNF clean room. We used a Deep Reactive Ion Etch (DRIE) tool to etch through the silicon. To make the hole pattern, we spin thick resist (SPR 220-7) and expose and develop to create an opening in the resist where the hole is to be made. We pattern the through holes on the ``cavity" side of the wafer. This avoids the possibility of overetching artifacts that might enlarge the hole at the exit point and produce bulk $^3$He in the cavity. To harden the resist further so it can survive the subsequent plasma etching, we leave the wafers in a 90 $^{\circ}$C convection oven overnight. The resist is also spun on the backside of the wafer to define the concentric grooves that are required to anchor the fill line. The resist is exposed and developed in a similar way, and the wafer is exposed to a timed DRIE exposure to make the 20 $\mu$m grooves illustrated in Fig. 2d and Fig. 3. At this point in the process flow, there is oxide on both sides of the silicon wafer.

In the next step we etch away (using RIE) the oxide in the hole area from the cavity side while preserving the oxide on the other side (outer surface of the wafer). The oxide on the outside surface will serve as a stop for the plasma once the hole in the silicon is completed. The remaining resist and the oxide outside of the hole area provides a mask through which the wafer can be etched through DRIE (the Bosch DRIE process has a high selectivity for silicon versus resist, and even higher versus oxide). The DRIE process deposits fluoropolymer on the sidewalls of the hole. To remove this a further RIE plasma etch step of CF$_4$ + O$_2$ was performed. To ensure full removal of the Bosch polymer, and to preserve the cleanliness of the pieces until just before bonding, the wafers were oxidized one more time at 1100 $^{\circ}$C for a short time and 200-300 nm of oxide was grown. Then just before bonding that oxide was removed using a 49\% HF dip. Any residue that could be on the wafer surfaces prior to this step is oxidized in the furnace and underetched in the HF dip. At the end of this step the wafer is at the point illustrated in Fig. 2e. An optical microscope image of the backside of the completed fill line hole is shown in Fig. 3. Finally the wafer is prepared for bonding by breaking the individual cavities free from the wafer at their ``tabs"  taking the process to the point illustrated in Fig. 2f.

\begin{figure}[H]
\begin{center}
    \includegraphics [angle=0,width=13cm]{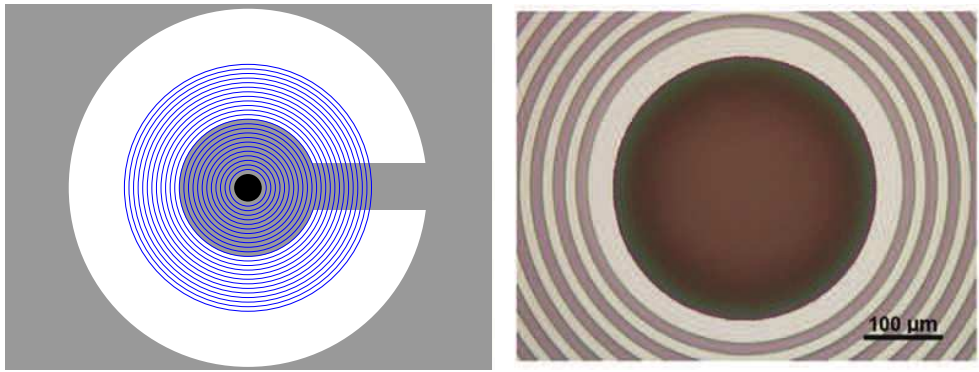}
    \caption{ (Left) An overview of the circular pattern etched onto the backside of the TO oscillator silicon wafer. The central black region represents the through hole, white areas represent regions where there is no etch and blue areas where an etch to 20 $\mu$m depth was carried out. The lighter grey area is the etched region on the front side of the wafer where the $^3$He is confined. (Right) Optical micrograph of the concentric pattern showing the through hole $\sim$350 $\mu$m diameter, and concentric circles having a 40 $\mu$m periodicity. The circles were etched to provide an increased surface for epoxy at the silicon-to-metal joint.}
    \label{figure3}
\end{center}
    \end{figure}

\subsection{Polishing and dicing of glass}

In contrast to the patterning of the silicon, the glass needs no preparation steps involving lithography. However prior to dicing, the Hoya SD-2 glass wafer\cite{hoya} has to be polished to achieve the required smoothness using a standard Chemical-Mechanical Planarization (CMP) polish process. In order to observe phase transitions whose interface can freely traverse the cell we need smooth cell surfaces. Scratches tend to pin the phase boundaries \cite{levitin2} and broaden the phase transitions. Thus it is important to have a feature-free cell and cell walls\cite{zhelev}. In the past, the glass surfaces were known to have surface scratches. To remove any scratches, the glass was polished in a CMP process using Cabot SS12 slurry, containing 100 nm nominal size silica grit suspended in a KOH solution. A new polishing pad (Rodel IC1400) was also used for the polishing step. The tool parameters were 7 lbs downforce, 30 RPM on the chuck, 40 RPM on the table, while flowing 150-200 ml/min of slurry onto the table. No further polishing was needed (or possible) for the silicon.

After polishing, the next step was to dice the glass in the octagonal shape needed for the TO and a rectangular shape for the NMR cells. We spun a thick layer of SPR220-7 resist which was subsequently baked overnight in a 90$^{\circ}$C convection oven to harden the resist further. The hardened resist is thought to be more effective as a protective layer during the dicing process. The wafer was then put on a sticky dicing tape and diced with a diamond coated dicing saw tool in the clean room. For the TO, the wafer was first diced into squares, and then the squares were put in the tool one by one and had their four corners cut off to produce an octagon.

\subsection{Cell Anodic Bonding}

At this point in the assembly process stream, the silicon and glass pieces were ready to bond\cite{Pomerantz,Anthony,Hui} in order to finish the assembly of the cell. Unlike the procedure described in Ref.  \cite{dimov}, the cells were individually bonded. We soaked the silicon in hydrofluoric acid (HF) to remove the oxide on the wafer chip surface. This step should have under-etched any contamination on the silicon surface. However, we could not do the same with the glass pieces, since HF aggressively etches glass and would roughen the surface. Instead a half hour long basic RCA SC-1 clean recipe was used to clean the glass surface. AFM scans after the RCA SC-1 clean show no significant roughening due to the clean. While this step was not necessary for the silicon, we put the silicon pieces in the solution as well in order to clean any possible contamination they might have picked up after the HF bath.

The assembly was completed using a custom made bonding jig shown in Figure 4. This procedure was carried out in the clean room. A macor spacer with a machined hole that matched the cell's dimensions was placed in contact with the bottom electrode. The silicon piece was first placed in the macor holder cavity side up, and then the glass lid was gently laid on top of it. This macor spacer ensured that the silicon and glass stayed aligned and did not move laterally during the jig assembly and bonding process. After the glass was placed in contact with the silicon the top stainless steel electrode was laid on top of the glass. The top clamping steel ring separated by another macor spacer to isolate the clamp ring from the electrode completed the assembly. The steel ring was connected to the bottom electrode through cap screws which were evenly tightened to apply pressure on the glass-silicon stack.

\begin{figure}
\begin{center}
    \includegraphics [angle=0,width=12cm]{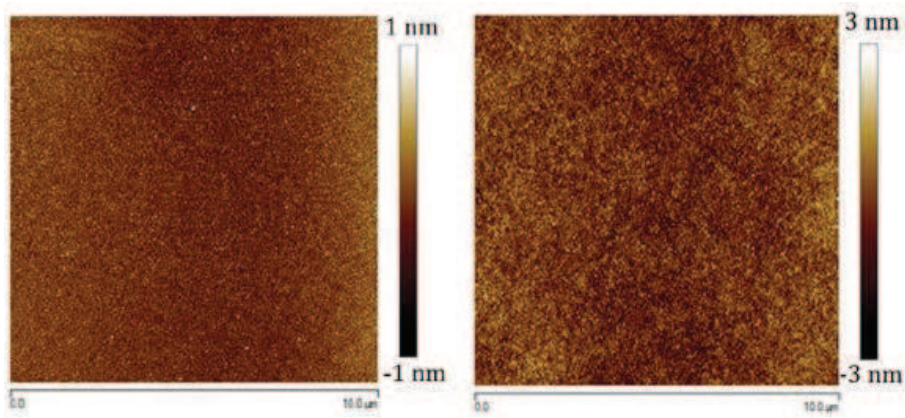}
    \caption{ Schematic of the custom made bonding jig. Grey represents stainless steel and white shows the macor spacer. The jig is placed in an ambient air furnace during the bonding process.}
    \label{figure4}
\end{center}
    \end{figure}

\begin{figure}
\begin{center}
    \includegraphics [angle=0,width=12cm]{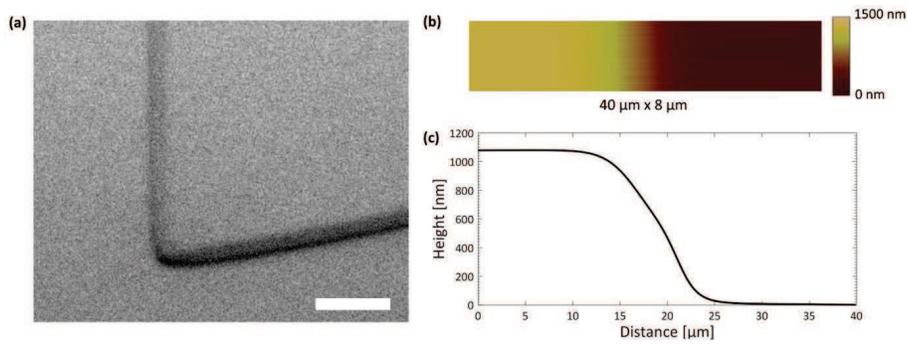}
    \caption{ (a) The torsion pendulum head in which the fluid is confined is made by bonding a patterned disk-shaped piece of silicon and an octagonal piece of smooth glass. (b) The bonded TO cell. (c) equivalent construction to (a) and (b) shows a bonded NMR cell. The cavity in which the $^3$He fluid is to be contained can be seen easily due to interference under ambient lighting.}
    \label{figure5}
\end{center}
    \end{figure}

The bonding jig was placed in an air furnace and heated to 425$^{\circ}$C. The two electrodes were connected to a high voltage power supply with polarity shown in Fig. 4 and 200 V was applied for about 5 minutes. This period of time was sufficient to create a reliable bond between the silicon and the glass. The glass is sodium doped; when voltage is applied at high temperatures sodium ions migrate from the glass-silicon interface. This results in dangling bonds at the glass surface which then connect to the silicon atoms on the silicon surface.  The temperature of the bonding process was chosen so that the total thermal expansion between room temperature and the bonding temperature is the same for both the glass and the silicon. This results in no bowing or distortion of the cell as it cools down to room temperature. A small degree of bowing of order 10 nm will occur, however, when cell is cooled from 300K to cryogenic temperatures. Images of the completed cells are shown in Figure 5.

\subsection{Metal Deposition}

The final step in the fabrication was to deposit a $\sim$1 $\mu$m thick silver film layer on the outside surfaces of the cell. The silver film aids in the thermalization of the silicon and the glass to the electron temperature of the nuclear cooling stage. The fill line hole was temporarily sealed with a piece of kapton tape. A second small piece of kapton tape (approximately 2 mm on a side) was placed on the outer surface of glass side of the cell to cover the  fill line hole so as to not obstruct the interior view of the fill line. This allowed us to verify that epoxy did not run into the cell when attaching the cell to the fill line. A teflon holder to clamp the cell by points on its edge was machined. The holder with head was attached to a rotation plate inside a sputter deposition chamber so that the TO cell could be silvered on all sides. The NMR cell was only silvered on its faces (the edges were not silvered to reduce eddy current heating from pulses). In order for the deposited silver film to adhere well to the silicon and the glass on the TO, a thin titanium adhesion layer was deposited first. Since titanium becomes superconducting at low temperatures decreasing its thermal conductivity significantly, we deposited only a 5-10 nm thick adhesion layer. The titanium layer was not used on the NMR cell and no problems with adhesion were noted.

\subsection{Summary of process flow}

The process flow described in this section is summarized here in point form. \\
1.	Design and prepare masks with the desired patterns.\\
2.	Grow thick ($\sim$2.5-3 $\mu$m) thermal oxide.\\
3.	Deposit more oxide ($\sim$2 $\mu$m) by PECVD.\\
4.	Pattern the cavity region using photolithography. Etch the oxide in the cavity region through a photoresist mask using dry and wet etching. \\
5.  Remove the photoresist in hot solvent bath. Clean the wafers in RCA SC-1 and SC-2 baths.\\
6.	Thermally oxidize the wafers. This defines the cavity depth.\\
7.	Deep Reactive Ion Etch (Bosch process) is used to etch a series of concentric circles around the fill line.\\
8.	Deep Reactive Ion Etch through the wafer to create the fill line.\\
9.  Remove the photoresist in hot solvent bath. To aid in the removal of the Bosch polymer that forms on the sidewalls during the Deep Reactive Ion Etch process, we plasma etch the resist stripped wafer in CF$_{4}$+O$_{2}$ plasma.\\
10.	Remove all oxide using wet etching by HF. To ensure full removal of the Bosch polymer, the wafers were additionally cleaned in RCA SC-1 and SC-2 baths and further oxidized for additional 300 nm of oxide on the surface. \\
11.    The glass wafer is polished. The glass is diced.\\
12.    Si wafers are broken along the tabs separating the individual cells. The Si pieces are dipped in HF to remove the protective oxide.\\
13.	The Si and glass pieces are cleaned in RCA SC-1 bath and put in contact in the bonding jig. The pieces are subsequently anodically bonded.\\
14.	A 1 $\mu$m thick silver layer for thermalization is sputter coated (over 5-10 nm Ti adhesion layer on the TO cell only) on the exterior of the cavity. The silver film did not cover the edges of the NMR cell \\
%15.	The fill line and the torsion rod are glued to the cell.\\

The cell is now ready for the fill line or torsion rod to be attached to the backside.

\section{Fill line epoxy seal}

The seal made between silicon and coin-silver has to survive stresses arising from the differential contraction between coin-silver, silicon and any epoxy used to connect the two materials. Most epoxies have a thermal expansion coefficient that is about twice as large as that of silver. We found that Tra-bond 2151 had the lowest thermal expansion coefficient of all tested epoxies and it proved to provide excellent adhesion. It also had large enough viscosity so it would not flow into the fill line and contaminate the cavity. Additionally in order that the TO have a high quality factor ($Q$), the contact area of the coin-silver to silicon joint should be reasonably large so that the dissipation arising from two level systems \cite{fefferman} should not contribute too much to the overall dissipation ($Q^{-1}$) of the TO. To maximize the contact area while limiting the stresses due to the differential expansion of the silicon and the silver, the TO torsion rod was designed to contact the silicon over a 2 mm diameter. We found that circles etched into the backside of the wafer provided enough surface area for contact so that the observed $Q$ was in excess of $10^6$ at low temperatures, more than an order of magnitude higher than that observed with the previous design \cite{dimov,dimov2}. If the silicon surface was just roughened (with no etched circles), 2 mm diameter epoxy joints tended to fail by delamination when cooled over a few minute time scale to liquid nitrogen temperature. The circles thus provide mechanical strength against delamination. For larger circular areas, the differential contraction produces stresses that result in the silicon cracking.  The process that was followed to join coin-silver to silicon is described below.

To act as a filler and decrease the volume of bulk fluid in the fill line next to the experimental cell, we selected a hollow silica tube (inner diameter 100 $\mu$m, outer diameter 320 $\mu$m)\cite{polymicro}. The tubing as supplied was coated with a polyamide layer which was removed by applying a lighted match to the tubing followed by a wipe with acetone and isopropyl alcohol. The tube was then cleaved so that its length was approximately 4 mm. The tube was then positioned using a 3-axis translation stage platform mounted to the base of a precision drill press. The glass tube was held in a teflon sleeve inserted into the drill chuck which was used to squeeze the teflon and securely hold the tube. The extent of compression was adjusted so as to allow the tube to slide vertically under a small applied force. A stereo microscope was focused on the cell hole and was used to align the glass tube to the fill line hole. After the alignment was verified by dry fitting, a 0.5 mm length of exposed silica tube (leaving the $\sim$0.5 mm long tip pristine) was coated with a bead of epoxy. The silica tube with the bead of epoxy was then inserted into the hole of the silicon cell. The fit is sufficiently tight and the epoxy's viscosity is high enough so that the epoxy is excluded from the cell. However, the epoxy volume is sufficient to flow to the silicon around the tubing to a radius of approximately 0.7 mm. After the epoxy cured the chuck was carefully loosened and decoupled from the glass tube. The base of the torsion rod was then inserted into the chuck. Fresh epoxy was put around the area of the silica tube - silicon surface joint and the 0.35 mm inner diameter torsion rod was carefully lowered onto the silicon squeezing out the freshly applied epoxy. Some of the epoxy fills the space between the silver and the silica tube, but the majority of the excess epoxy was extruded from the perimeter of the joint. The excess epoxy was wiped off and the joint was left to cure. Fig. 6 provides a schematic for the geometry of the joint, and illustrates the two step process involved. For NMR cells, a similar process involved a silver ferrule to cap the silica fill line.  The thus fabricated joint has minimal bulk volume contribution to the fluid inertia of  a TO and also minimal ``bulk" volume to interfere with the NMR signal since with the exception of the $^3$He in the 1 mm long, 100 $\mu$m diameter silica tube, bulk helium is shielded from RF by the silver fill line ferrule.

\begin{figure}[H]
\begin{center}
    \includegraphics [angle=0,width=5cm]{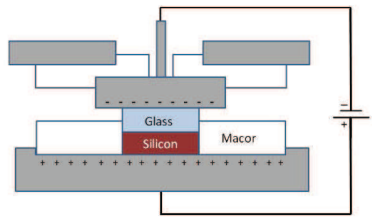}
    \caption{Schematic illustration of the two step process used to assemble the torsion rod and cell.}
    \label{figure6}
\end{center}
    \end{figure}

\section{Cell Surface Characterization}
\subsection{Roughness}

\begin{figure}[H]
\begin{center}
    \includegraphics [angle=0,width=15cm]{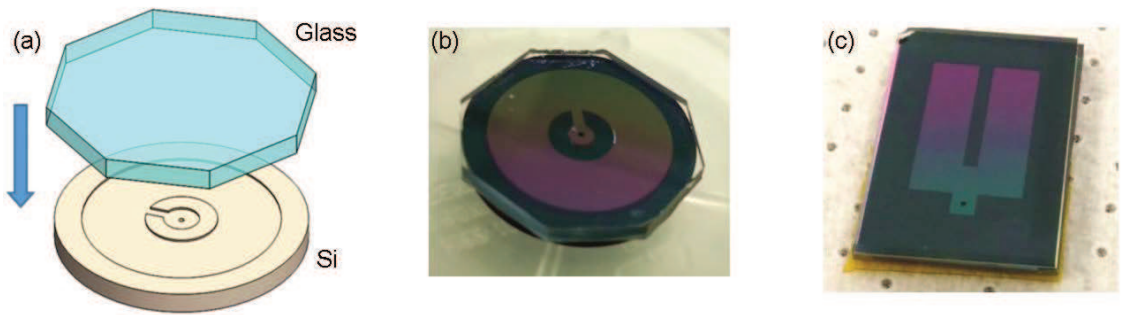}
    \caption{ AFM images of the cell surfaces. Images were taken before bonding. (a) Silicon surface showing 0.102$\pm$0.035 nm $R_a$ roughness (b) Glass surface showing 0.342$\pm$0.049 nm $R_a$ roughness. }
    \label{figure7}
\end{center}
    \end{figure}

After surfaces of the glass were polished, the Si and glass were characterized by scanning samples with an Atomic Force Microscope (AFM). From the image shown in Figure 7, we see that the surface is found to be nearly smooth with a surface roughness 0.102$\pm$0.035 nm for the Si (Figure 7a) and 0.342$\pm$0.049 nm (arithmetic average, $R_{a}$, and s.d) for the glass (Figure 7b). The glass surface scanned displays no scratches after polishing for several samples. The surface roughness was also seen to be about a factor of two less than that seen for unpolished Hoya glass\cite{dimov}. Experimental observations of the superfluid A to B transition (on cooling) and the B to A transition (on warming) from cells prepared using the recipe detailed in the previous section reveal good reversibility of the transition on warming once an A-B superfluid interface is formed.\cite{zhelev}. This is in contrast to the hysteresis seen in a cell fabricated without the additional polishing steps. \cite{levitin2}

\subsection{Scan of Edge}

The etch depth in the silicon cavity prior to bonding (but after dicing) was measured using profilometry. The profilometer read a step height of 1080 nm. A scanned electron micrograph of the cell near one of the cavity edges is shown in Fig. 8a. The step profile was also imaged using an AFM (Fig. 8b) and shows a step of $\sim$1.08  $\mu$m. The profile at the cell edge is not abrupt. Rather it shows a variation of height over a lateral length scale of $\sim$ 10 $\mu$m as shown in Fig. 8 due to lateral diffusion during the oxidation steps. This edge profile is significant since the effective confinement of the liquid $^3$He ($\xi_0$/D) is significantly increased in the region of the cell wall. ($D$ is the cavity height, and $\xi_0$ is the zero temperature coherence length.)

\begin{figure}[H]
\begin{center}
    \includegraphics [angle=0,width=15cm]{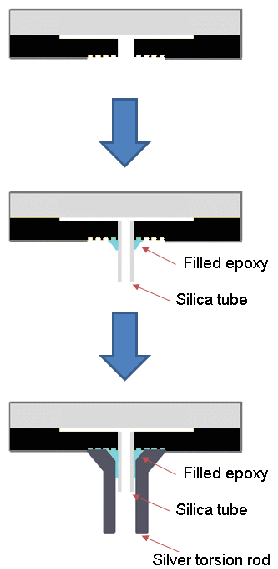}
    \caption{(a) Scanning electron micrograph showing the inner cavity wall and the edge of the central channel of a torsion oscillator cell (the scale bar is 40 $\mu$m). The walls of the cavity are rounded due to the oxidation process with a lateral extent of approximately 10 $\mu$m. (b)	Atomic force microscope image of step at the outside cavity wall. The scanned region has a length of  40 $\mu$m $\times$  8 $\mu$m width. (c)	Line slice through the AFM scan clearly showing the profile of the cavity wall. Evident from this plot is the size of cavity of 1080 nm and the extent of its variation. Reproduced with permission from Nature Communications, {\bf 8} 15963 (2017). Copyright 2017 Nature Publications under CC BY 4.0.}
    \label{figure8}
\end{center}
    \end{figure}

\section{Pressure testing and finite method simulation}

The NMR cell variant was pressure tested at room temperature. The cell was pressurized in 2 bar steps to 32 bar where the glass eventually gave way (Figure 9) while the bond survived. The height of the cavity was measured at several pressures and found to be closely in agreement (within 10\%) of the calculated maximum bowing using a COMSOL modelling of the structure. The cells were not pressure tested to failure at low temperatures. However they proved to be leak tight to 6 bar\cite{levitin1,zhelev}. The cells were cooled to liquid nitrogen temperature while mounted on the nuclear stage over a 1 day time scale, and to liquid helium temperatures in a few hours. Thus they were not subjected to extreme thermal shock.

\begin{figure}[H]
\begin{center}
    \includegraphics [angle=0,width=9cm]{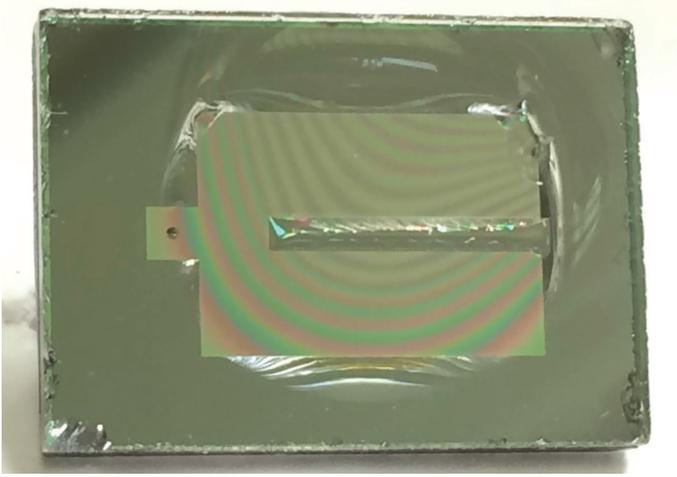}
    \caption{The image shows a cavity suitable for NMR tested (at room temperature) to 32 bar where the glass delaminated internally. Note the silicon-glass bond survives largely intact with the septum area showing cracks in the glass.}
    \label{figure9}
\end{center}
    \end{figure}

We used a finite-element method (FEM) to compute the bowing of a nanofluidic cavity caused by an applied fluid pressure $P_0$. For this calculation, we used COMSOL Multiphysics v5.2 and the Solid Mechanics module. The simulation was performed for the NMR cell with the dimensions provided in section II. The simulation results are illustrated in figure 10. In the experiment, a metallic ferrule was mounted around the fill line hole of the nanofluidic geometry. However, this piece does not strongly affect the results of the FEM simulation on the bowing of the cell. To save computation time, we did not include this part in the computation.

\begin{figure}[h]
\begin{center}
    \includegraphics [width=16cm]{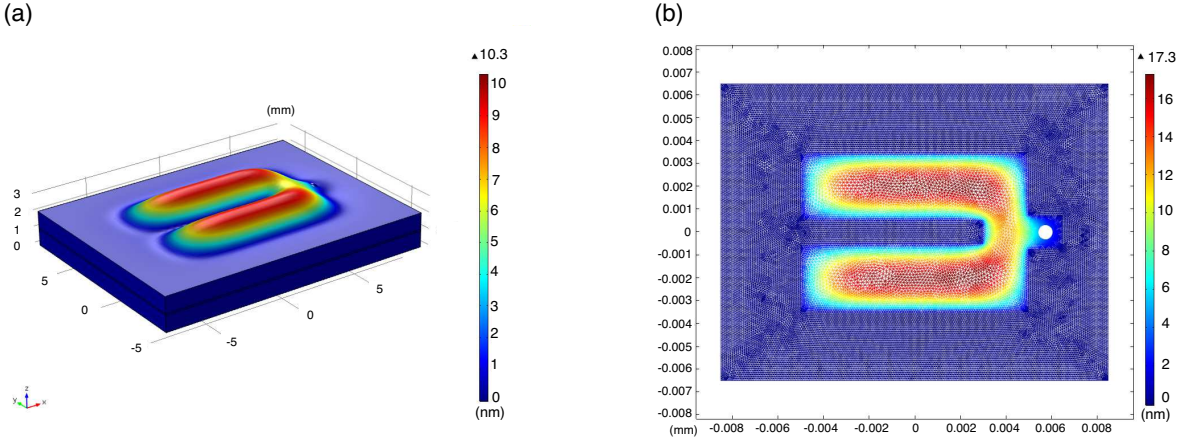}
    \caption{FEM simulation of the bowing of the nanofluidic cavity of  the NMR cell for an applied fluid pressure $P_0=1$ bar. (a) 3D view of the SD-2 glass lid deformation. The deformation has been exaggerated with a large scaling factor ($\times10^5$). (b) 2D view of the change in height of the nanofluidic cavity.}
    \label{figure10}
\end{center}
    \end{figure}

The boundary conditions are defined such that all surfaces are free except the lateral surface of the cell located on the perimeter. On this surface, the displacement field is fixed to zero ($u_x=u_y=u_z=0$). In the simulation, we produced the equivalent of a hydrostatic pressure inside the nanofluidic cavity by applying a boundary load ({\it i.e.} pressure) of magnitude $P_0=1$ bar on the inner surfaces of the nanofluidic cavity.

In the assembled experiment, the ``lid" of the nanofluidic devices is made from amorphous glass and the micromachined bottom part is a piece of single crystal silicon wafer (100 plane at the cavity surface). Nevertheless, for the simulation we assume isotropic mechanical properties for both sides. We used the material properties of polycrystalline silicon for the bottom part of the cell and the material properties given by Hoya SD-2 product literature for the top part. The relevant material properties used in the simulation are the density, Young's modulus and Poisson's ratio. For silicon, we used a value for the density $\rho_{\rm silicon}=2329$ kg/m$^3$, a value for the Young's modulus $E_{\rm silicon}=170$ GPa and the value for the Poisson's ratio $\nu_{\rm silicon}=0.28$. For the Hoya SD-2 glass, we used a value of the density $\rho_{\rm glass}=2600$ kg/m$^3$, a value of the Young's modulus $E_{\rm glass}=88.6$ GPa and a value of the Poisson's ratio $\nu_{\rm glass}=0.244$.

In FEM simulations, one has to define the mesh geometry; we used tetrahedrons as mesh elements. We observed slight differences in the results as a function of the mesh element size. Eventually, the results should converge to a certain value when the mesh is sufficiently fine. The results reported here have been obtained with the finest meshing and should be within $10\%$ of this convergence value.

The maximum change in height for the NMR cell is 17.3 nm/bar in total with a contribution of 7.0 nm/bar from the silicon wafer and 10.3 nm/bar from the silica glass wafer. These results are illustrated in Figure 10. The corresponding results for the torsion pendulum cell are discussed in a previous publication\cite{zhelev}.

\section{Summary}

We have demonstrated the ability to fabricate 1080 nm deep silicon-glass cavities with no internal supports. The use of anodic bonding of glass to silicon allows us to optically characterize the cavities after assembly. The well characterized glass and silicon surfaces fabricated using the processes detailed in this paper are demonstrated to be smooth enough not to pin the interfaces between the A and B phases of $^3$He. Silicon - silver - epoxy joints also have been shown to survive multiple coolings to cryogenic temperatures while also demonstrating a high $Q$. The design of the fill line also minimizes the volume of bulk fluid in the fill line and shields the fill line fluid from RF simplifying NMR signal identification. These features are important enabling factors to be built on for future designs. New designs that span the parameter space from several $\mu$m down to the 100 nm scale will be possible, some with internal supports to limit bowing and prevent collapse while bonding. These designs could incorporate more complex structures including height steps and constrictions to allow the observation of mass and thermal flow across the nearly 3D to quasi 2D limit in superfluid $^3$He experiments. Aspects of these cells' construction (smoothness of surfaces, optical access through the glass lid to visualize flows under pressure, pressure capability at room temperature and down to low temperatures, and the fluid port construction) may well find applications in mainstream nanofluidics.

\section{Acknowledgements}

We acknowledge the support of the staff and use of the facilities of the Cornell NanoScale Science and Technology Facility.
We acknowledge the financial support from the NSF under DMR1202991, DMR1664043, DMR1708341 at Cornell and at Royal Holloway from the EPSRC under EP/J022004/1 and the European Microkelvin Platform.

\end{document}